\documentclass[floats,floatfix,aps,showpacs,twocolumn,pre]{revtex4-1}

\usepackage[dvips]{graphicx}
\usepackage{amsmath}   % for \text
\usepackage{amsfonts}
\usepackage{subfigure}
\usepackage{braket}
\usepackage[utf8]{inputenc}
\usepackage{nicefrac}

\newcommand{\Int}{\int\limits}

\newcommand{\heff}{h_{\text{eff}}}
\newcommand{\heffbar}{\hbar_{\text{eff}}}
\newcommand{\Areg}{A_{\text{reg}}}
\newcommand{\Dreg}{\Delta_{\text{reg}}}
\newcommand{\Dcha}{\Delta_{\text{ch}}}
\newcommand{\Ncha}{{N_{\text{ch}}}}

\newcommand{\Hreg}{H_{\text{reg}}}
\newcommand{\Hcha}{H_{\text{ch}}}

\newcommand{\tH}{\tau_{\text{\tiny{H,ch}}}}

\newcommand{\mmax}{m_{\text{max}}}

\newcommand{\Z}{\mathbb{Z}}

\newcommand{\ud}{\text{d}}
\newcommand{\ue}{\text{e}}
\newcommand{\ui}{\text{i}}

\newcommand{\bracket}[2]{\Braket{ #1  | #2 }}

\newcommand{\chireg}{\chi^\text{reg}}
\newcommand{\chiregnm}{{\chi^\text{reg}_{n,m}}}
\newcommand{\chich}{\chi^\text{ch}}

\newcommand{\Wm}{f}

\newcommand{\Pregmnonav}{p_{m,\theta_q,\ket{\varphi_0}}^\text{reg}}
\newcommand{\Pregnonav}{p_{\ket{\varphi_0}}^\text{reg}}
\newcommand{\Pregm}{p_m^\text{reg}}
\newcommand{\Preg}[1]{p_{#1}^\text{reg}}
\newcommand{\Pfluct}{p_{\text{fl}}}

\newcommand{\Pch}{p^\text{ch}}
\newcommand{\tauH}{\tH}

\newcommand{\tplateau}{t_{\rm sat}}
\newcommand{\tplateaum}{{t_{{\rm sat}, m}}}

\newcommand{\veff}{v_\text{eff}}
\newcommand{\veffm}{v_{\text{eff},m}}

\newcommand{\satur}{f^{\infty}}
\newcommand{\saturm}{f^{\infty}_m}
\newcommand{\saturph}{f^{\infty}_{2\times 2}}

\newcommand{\fourvec}[4]{\left(\begin{array}{c}#1\\#2\\#3\\#4\end{array}\right)}

\newcommand{\vregchm}{v^{\text{reg},\text{ch}}_m}
\newcommand{\Deff}{\Delta_{\text{eff}}}

\newcommand{\regweight}{regular weight}
\newcommand{\regweights}{regular weights}

\newcommand{\Regweights}{Regular weights}

\newcommand{\floodweight}{flooding weight}
\newcommand{\floodweights}{flooding weights}
\newcommand{\Floodweight}{Flooding weight}
\newcommand{\Floodweights}{Flooding weights}

\newcommand{\satval}{asymptotic flooding weight}
\newcommand{\satvals}{asymptotic flooding weights}
\newcommand{\Satval}{Asymptotic flooding weight}
\newcommand{\Satvals}{Asymptotic flooding weights}

\renewcommand{\nicefrac}[2]{ #1  / #2 }

\begin{document}
 \title{Temporal flooding of regular islands by chaotic wave packets}
 \author{Lars Bittrich$^{1,2}$, Arnd B\"acker$^{1,2}$, and Roland Ketzmerick$^{1,2}$}
  \affiliation{$^1$Technische
               Universit\"at Dresden, Institut f\"ur Theoretische Physik
               and Center for Dynamics, 01062 Dresden, Germany\\
               $^2$Max-Planck-Institut f\"ur Physik komplexer Systeme,
               N\"othnitzer Stra{\ss}e 38, 01187 Dresden, Germany}

\date{22.01.2014}

\begin{abstract}
 We investigate the time evolution of wave packets in systems with a mixed phase
space where regular islands and chaotic motion coexist.
For wave packets started in the chaotic sea on average the weight on a quantized
torus of the regular island increases due to dynamical tunneling.
This flooding weight initially increases linearly and saturates to a value which
varies from torus to torus.
We demonstrate for the asymptotic flooding weight universal scaling with an
effective tunneling coupling for quantum maps and the mushroom billiard. This
universality is reproduced by a suitable random matrix model.
\end{abstract}

\maketitle

%%%%%%%%%%%%%%%%%%%%%%%%%%%%%%%%%%%%%%%%%%%%%%%%%%%%%%%%%%%%%%%%%%%%%%%%%
\section{Introduction}
%%%%%%%%%%%%%%%%%%%%%%%%%%%%%%%%%%%%%%%%%%%%%%%%%%%%%%%%%%%%%%%%%%%%%%%%%

Typical Hamiltonian systems have a mixed phase space in
which dynamically separated
regular and chaotic regions coexist. A fundamental question is how the
properties of the classical dynamics are reflected in the corresponding quantum
system \cite{Sto2000,Haa2010}.
In the semiclassical limit, i.e.\ at short wave length or when typical
actions become large in comparison to Planck's constant, one expects
according to the semiclassical eigenfunction hypothesis
that eigenstates localize on classically invariant regions in phase space
\cite{Per73, Ber1977b, Vor79}. Thus they can be classified as regular or chaotic,
see e.\ g.\ \cite{BohTomUll1993,ProRob93f,LiRob1995b,CarVerFen1998,VebRobLiu1999,BaeKetLoeSch2011}.
Away from the semiclassical limit, however, the correspondence between eigenstates
and classically invariant regions breaks down.
For example partial barriers can lead to localization
of eigenstates and wave packets
\cite{KayMeiPer1984b,BroWya1986,GeiRadRub1986,BohTomUll1993,MaiHel2000,MicBaeKetStoTom2012}.
Another phenomenon is dynamical tunneling \cite{DavHel1981,KesSch2011}
between regular and chaotic phase-space regions.
As a consequence regular states only exist if in addition to the WKB-quantization condition the relation \cite{BaeKetMon2005,BaeKetMon2007}
\begin{equation}
\label{eq:existence_intro}
 \gamma_m < \frac{1}{\tauH}\;\;
\end{equation}
is fulfilled, where $\gamma_m$ is the regular-to-chaotic tunneling rate
from the $m$th torus to the chaotic sea
and $\tauH=\nicefrac{\heff}{\Dcha}$ is the Heisenberg time of the chaotic sea
with mean level spacing $\Dcha$. If the criterion \eqref{eq:existence_intro} is
violated the $m$th regular state disappears
and the corresponding region in phase space is flooded by chaotic states.
The transition region, until the state fully disappears,
is rather broad.
An important consequence are huge localization lengths
in nano wires with surface
disorder \cite{FeiBaeKetRotHucBur2006,FeiBaeKetBurRot2009}.
Also eigenstates in higher-dimensional systems are influenced
by flooding \cite{IshTanShu2010}.

\begin{figure}[b]
 \vspace*{-1.1ex}  % this puts the "Sec. VI" on the first page.
 \centering
 \includegraphics{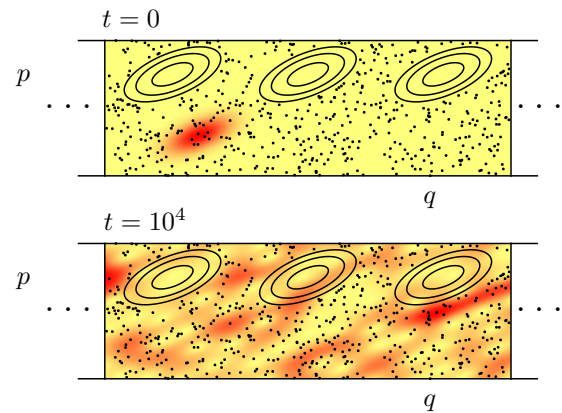}
 \caption{(Color online)
 Flooding in a system with a mixed phase space. A wave packet started
 at $t=0$ in the chaotic sea (dots) floods the regular island (closed lines)
 for large times.\label{fig:amphibabb_intro}}
\end{figure}

Flooding also occurs in the time
evolution of wave packets. Starting a wave packet in the chaotic
sea it will partially or completely flood the regular island
\cite{BaeKetMon2005}, see Fig.~\ref{fig:amphibabb_intro}.
The determination of tunneling rates $\gamma_m$
entering in Eq.~(\ref{eq:existence_intro}) has been
studied in much detail, see e.g.~\cite{LoeBaeKetSch2010,KesSch2011,MerLoeBaeKetShu2013},
including consequences on spectral statistics \cite{BaeKetLoeMer2011},
quality factors in optical microcavities \cite{BaeKetLoeWieHen2009},
and the existence of bouncing-ball modes \cite{LoeBaeKet2012}.

In this paper we study this temporal flooding of the regular island not just for the
entire regular island \cite{BaeKetMon2005}, but specifically
for individual tori. We quantify the amount of flooding of a torus by a suitably defined
weight. This flooding weight initially increases linearly. At a saturation time
it reaches its asymptotic value. We observe that the asymptotic flooding weight
and the saturation time for individual tori show a universal scaling with an effective tunneling
coupling. This is found for a suitably designed quantum map as well as the generic
standard map and the mushroom billiard. The universality is reproduced by an
appropriate random matrix model.

This paper is organized as follows: Temporal flooding is introduced for
the designed quantum map in Sec.~\ref{sec:temp_flood}. The initial linear increase of the
flooding weight is explained in Sec.~\ref{sec:linear_regime}. The following
Sec.~\ref{sec:saturation_regime} deals with the universal scaling of the
asymptotic flooding weight and its modelling by random matrices.
Further applications to the standard map and the mushroom billiard are presented in
Sec.~\ref{sec:applications}. A summary and outlook
is given in Sec.~\ref{sec:summary}.

%=======================================================================
\section{Temporal flooding} \label{sec:temp_flood}
%=======================================================================

\subsection{System}
\begin{figure}[t]
 \includegraphics[width=8.5cm]{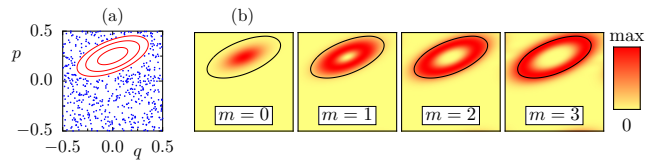}
 \caption{(Color online)
 (a) Classical phase space of the designed map with chaotic dynamics (blue dots)
and regular tori (red lines). (b)
 Husimi representation of all regular basis states $m=0, 1, 2, 3$ for
  $\heff=\nicefrac{1}{20}$ (b). The border of the island is indicated by a solid
  line.
 \label{fig:reghusimis}}
\end{figure}

For the study of temporal flooding in a mixed
phase space we consider an example system with one large regular island
surrounded by a chaotic sea \cite{HufKetOttSch2002, BaeKetMon2007}. This is
realized by a kicked Hamiltonian system
\begin{equation}
  H(p,q,t) =  T(p) + V(q) \sum_{n= -\infty}^{\infty} \delta(t - n ) \ ,
   \label{eq:hamiltonian}
\end{equation}
where the potential $V(q)$ and the kinetic energy $T(p)$ are
designed appropriately with periodic
boundary conditions in $q$- and $p$-direction,
see Appendix~\ref{sec:amphib_map} for a detailed definition.
One considers the dynamics
stroboscopically just after each kick, giving an area preserving designed map,
see Fig.~\ref{fig:reghusimis}(a).

Quantum mechanically the dynamics of such a map
is described by a unitary operator $U$
(see e.g.~\cite{BerBalTabVor79,HanBer1980,ChaShi86,Esp93})
\begin{equation}
   U = \exp\left( -\frac{2 \pi \ui}{\heff} V( q ) \right)
           \exp\left( -\frac{2 \pi \ui}{\heff} T( p ) \right)
                             \ , \label{eq:propagator}
\end{equation}
which determines the time evolution of wave packets
\begin{equation}
\ket{\varphi_t} = U^t \ket{\varphi_0} \ , \ t= 0,1,2, ... \ .
\label{eq:U_eigeneq}
\end{equation}
The dimension of the Hilbert space is given by $N=\nicefrac{1}{\heff}$.
The eigenstates $\ket{\psi_j}$ of $U$ are defined by
\begin{equation}
 U\ket{\psi_j} = \ue^{\ui\varepsilon_j}\ket{\psi_j}\;\;,
\label{eq:U_eigeneq_psi}
\end{equation}
where $\varepsilon_j$ are the quasi-energies having mean spacing $2\pi/N$.
The eigenstates can be classified as either regular or chaotic depending on the
region on which they predominately concentrate.
Due to tunneling they have contributions in all regions of phase space.

A time-evolved wave packet, initially localized in the chaotic sea, will tunnel into the regular
island. To measure its
weight in the regular island we will later use its projection onto
\textit{regular basis states} $\ket{\chireg_m}$, which are concentrated on quantized regular
tori, see Fig.~\ref{fig:reghusimis}. These tori fulfill the WKB quantization rule
\begin{equation}
\label{eq:EBK}
 \oint p \, \ud q = \left(m+\frac{1}{2}\right) \heff \ ,
\end{equation}
with quantum number $m=0, 1, ..., \mmax-1$.
The number of regular basis states in the island of area $\Areg$ is given by
\begin{equation}
\mmax = \left\lfloor \frac{\Areg}{\heff}+\frac{1}{2} \right\rfloor
\label{eq:mmax}\;\;,
\end{equation}
where $\lfloor \cdot \rfloor$ denotes the floor function.
We consider a designed map, Appendix~\ref{sec:amphib_map}, for which $\Areg \approx 0.21$.
For $\heff=\nicefrac{1}{20}$ this leads to $\mmax=4$ regular basis states.
Generally, the construction of the regular basis states $\ket{\chireg_m}$
can be done using semiclassical methods. For the designed map
they are given analytically, Eq.~\eqref{eq:tiltedstate}.
The regular basis states $\ket{\chireg_m}$ form an orthonormal basis within the
regular island, they have no chaotic admixture, in contrast to the
regular eigenstates of the quantum map.
\subsection{Wave packet dynamics}
\label{sub_sec:wave_dyn}
\begin{figure}[tb]
 \centering
 \includegraphics[width=8.5cm]{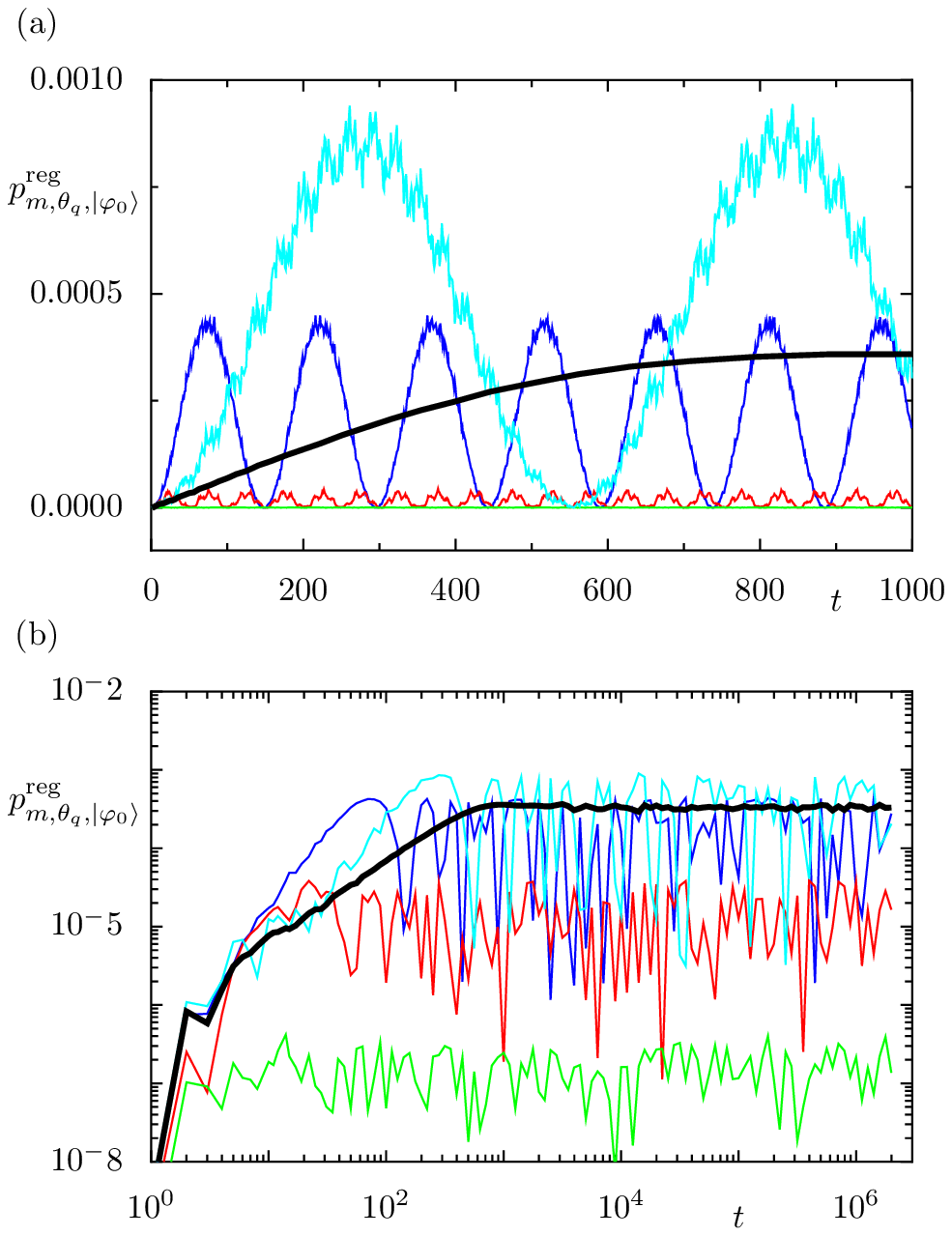}
 \caption{(Color online)
(a) The weight $\Pregmnonav$ according to Eq.~\eqref{eq:regular_weight}
  on the regular torus $m=0$ of the designed map with the initial state
  $\ket{\varphi_0}$ being a momentum eigenstate with $p \approx-0.25$ and under
  variation of the Bloch phase $\theta_q$ ($0.2, 0.3, 0.55, 0.75$,
  blue, red, green, cyan) for
  $\heff=\nicefrac{1}{20}$. The black line shows the \regweight{} $\Pregm$
  according to Eq.~\eqref{eq:regular_weight_av} averaged over $20000$ pairs
  $\theta_q, \ket{\varphi_0}$. Here $\theta_q\in[0,1)$ is chosen equidistantly
  and $\ket{\varphi_0}$ is a momentum eigenstate with $p\in[-0.45, -0.05]$
  chosen randomly.
  (b) Double-logarithmic representation of the same data.
 \label{fig:no_average_ebkweights}}
\end{figure}

We consider the behavior of wave packets $\ket{\varphi_t}$, which are initially
localized in the chaotic sea. Their weight on the $m$th regular torus is measured
by the overlap with the $m$th regular basis state $\ket{\chireg_m}$
\begin{equation}
  \Pregmnonav(t) = \left| \bracket{\chireg_{m}}{\varphi_t}\right|^2\;\;.
  \label{eq:regular_weight}
\end{equation}
This overlap depends on the initial wave packet $\ket{\varphi_0}$ and the
chosen Bloch phase $\theta_q$, which arises from the periodic boundary
conditions in $q$-direction.
Fig.~\ref{fig:no_average_ebkweights} shows the probability $\Pregmnonav(t)$
for $m=0$ and different $\theta_q$.
We observe Rabi-like oscillations with various amplitudes and frequencies.
They originate from the tunneling coupling of the regular basis state $\ket{\chireg_0}$ with
the spectrally closest chaotic state for each $\theta_q$. These are superimposed
by small oscillations caused by couplings to other chaotic states. In
Fig.~\ref{fig:no_average_ebkweights}(b) we show $\Pregmnonav(t)$ on a double
logarithmic scale.

To investigate the universal behavior of the time evolution of wave packets
flooding the regular island, we first introduce the \textit{\regweight{}} $\Pregm$
\begin{equation}
  \Pregm(t) = \left\langle \Pregmnonav(t) \right\rangle_{\theta_q, \ket{\varphi_0}}\ ,
  \label{eq:regular_weight_av}
\end{equation}
which is the average over different Bloch phases $\theta_q$ and different initial
wave packets $\ket{\varphi_0}$. In Fig.~\ref{fig:no_average_ebkweights}(b)
the \regweight{} $\Preg{0}(t)$ shows a linear increase at small times and a
saturation plateau at large times.
\subsection{\Floodweight{}}
To quantify the flooding in systems with
a mixed phase space it is helpful to divide the Hilbert space into the regular
basis states $\ket{\chireg_m}$ and their complement containing the chaotic basis
states $\ket{\chich}$. This separation allows for defining the \textit{\floodweight{}}
$\Wm_m(t)$ which will turn out to show universal behavior,
\begin{equation}
  \Wm_m(t) = \frac{\Pregm(t)}{\frac{\Pch(t) + \Pregm(t)}{\Ncha +1}} \ .
\label{eq:flooding_weights}
\end{equation}
It is the ratio of the regular weight $\Pregm$ on the $m$th torus to the average
weight in the subspace given by the chaotic basis states and the $m$th regular
basis state.
Here $\Pch(t)$ is the weight in the chaotic subspace following from normalization
\begin{equation}
\Pch + \sum_{m=0}^{\mmax-1}  \Preg{m} = 1 \ .
\label{eq:normalized_preg}
\end{equation}
and $\Ncha$ is the number of chaotic basis states, $\Ncha = N - \mmax$. The
\floodweight{} $\Wm_m(t)$, Eq.~\eqref{eq:flooding_weights},
has two advantages compared to the \regweight{} $\Pregm(t)$,
Eq.~\eqref{eq:regular_weight_av}:
(i) The \floodweight{} $\Wm_m(t)$ reaches $\Wm_m = 1$ if the wave packet is
uniformly spread over the chaotic sea and the $m$th regular torus. This is
independent of the number of chaotic basis
states. (ii) The \floodweight{} $\Wm_m(t)$ does not depend on the flooding
of other regular basis states unlike the \regweights{} $\Pregm(t)$, which has
particular relevance if many tori are flooded, see
Fig.~\ref{fig:flooding_complete} in Appendix~\ref{app:master}.

Fig.~\ref{fig:flooding_weights_M1}(a) shows the \floodweights{} $\Wm_m(t)$
for different
regular basis states. Fig.~\ref{fig:flooding_weights_M1}(b)
shows a double-logarithmic representation of the same data
and we observe a linear increase followed by a saturation
of the \floodweights.
For regular tori closer to the center of the island the \satval{} is lower.
\begin{figure}[tb]
 \centering
 \includegraphics[width=8.5cm]{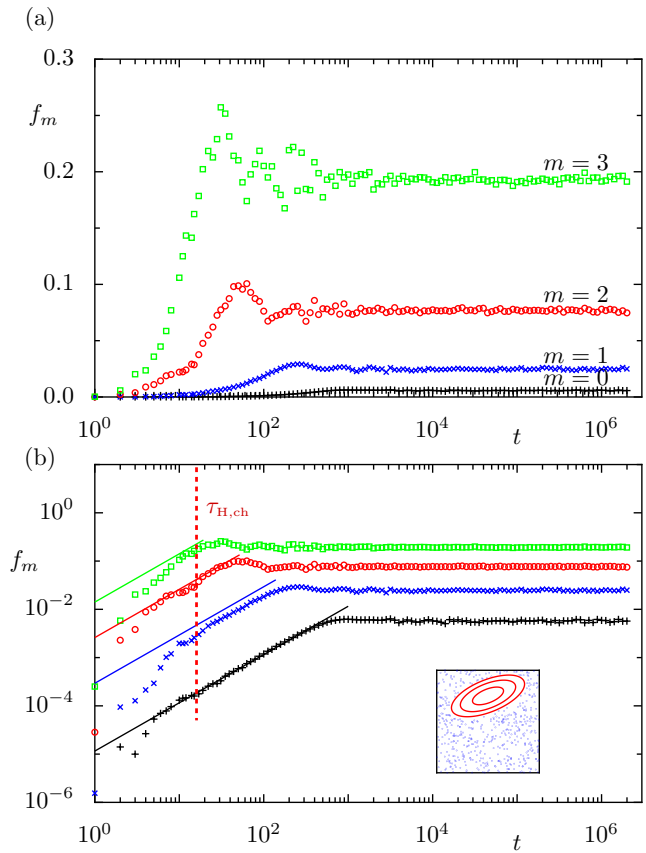}
 \caption{(Color online)
 (a) \Floodweights{} for a wave packet started in the chaotic
 sea of the designed map with $\heff=\nicefrac{1}{20}$ for all regular basis states $m=0,1,2,3$.
 (b) Double-logarithmic representation of the same data
 together with the predictions of
 Eq.~\eqref{eq:approximate_flooding_tunneling} (solid lines).
 The Heisenberg time $\tauH$ is indicated by a dashed line.
 Inset: Phase space for $M=1$.
 \label{fig:flooding_weights_M1}}
\end{figure}
The regime of linear increase and the saturation regime will be discussed in
Secs.~\ref{sec:linear_regime} and \ref{sec:saturation_regime} respectively.
The transition
regime shows in Fig.~\ref{fig:flooding_weights_M1}(a) remnants of Rabi-like
oscillations as for $\Pregmnonav$ in Fig.~\ref{fig:no_average_ebkweights}(a),
which will be discussed in Sec.~\ref{sec:full_time_solution}.

% #############################################################################
\subsection{Many unit cells in phase space}
\begin{figure}[b]
 \centering
 \includegraphics[width=8.5cm]{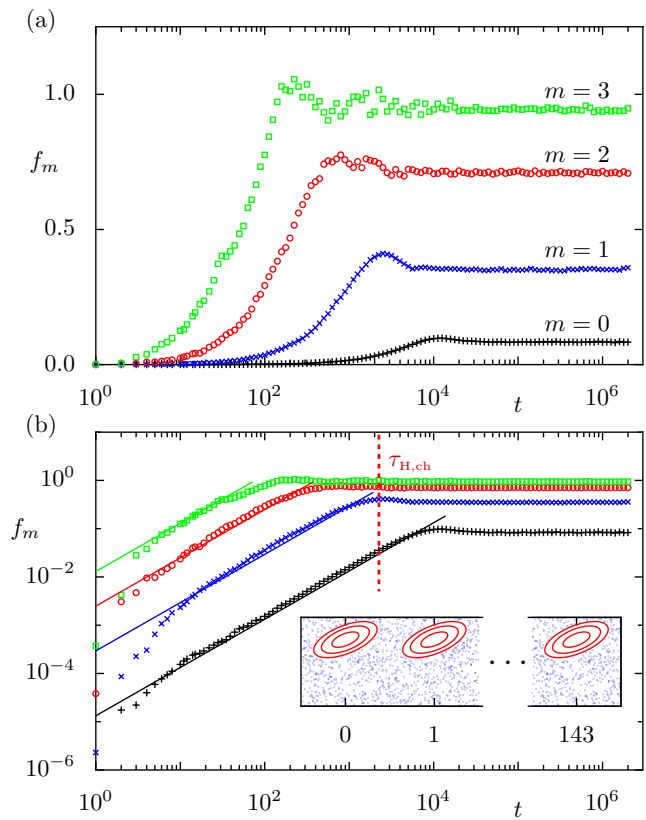}
 \caption{(Color online)
  (a) \Floodweights{} for a wave packet started in the chaotic
 sea of the designed map with $\heff=\nicefrac{144}{2825}\approx\nicefrac{1}{20}$ and $M=144$
 for all regular basis states $m=0,1,2,3$.
 (b) Double-logarithmic representation of the same data
 together with the predictions of
 Eq.~\eqref{eq:approximate_flooding_tunneling} (solid lines).
 The Heisenberg time $\tauH$ is indicated by a dashed line.
 \label{fig:flooding_weights}}
\end{figure}

The amount of flooding of regular tori depends both on the tunneling rate and
the size of the chaotic sea. A periodic extension of the designed map to $M$
unit cells in $q$-direction, see inset of
Fig.~\ref{fig:flooding_weights}(b), increases the size of the chaotic sea.
It thus increases the chaotic density of states and thereby the
Heisenberg time $\tauH$,
while the tunneling rates remain
unchanged for fixed $\heff$. Thus less and less regular eigenstates
per unit cell exist according
to Eq.~\eqref{eq:existence_intro} and we expect stronger temporal flooding
\cite{BaeKetMon2005, BaeKetMon2007} with \floodweights{} reaching $\Wm_m\approx1$.

The generalization of the designed map to $M$ unit cells in $q$-direction
(Appendix~\ref{sec:amphib_map}) leads to the following changes: (i) $\heff=M/N$,
where $M$ and $N$ have no common divisor to avoid periodicities.
(ii) We now use regular basis states $\chireg_{m,j}$, which localize on the
$m$th regular torus of the regular island of the $j$th unit cell. (iii) The \regweight{} is
defined by summation over all $M$ regular basis states with the same quantum number
$m$
\begin{equation}
  \Pregm(t) = \left\langle \sum_{j=0}^{M-1}\left| \bracket{\chireg_{m,j}}{\varphi_t}
  \right|^2\right\rangle_{\theta_q, \ket{\varphi_0}}\;\;.
  \label{eq:regular_weight_Mq}
\end{equation}
(iv) The \floodweight{} $\Wm_m(t)$ is now defined by
\begin{equation}
  \Wm_m(t) =\frac{\frac{\Pregm (t)}{M}}{\frac{\Pch(t) + \Preg{m}(t)}{\Ncha +M}} \; \; ,
\label{eq:flooding_weights_Mq}
\end{equation}
which is the ratio of the average weight in the subspace
of the $M$ regular basis states on the $m$th regular torus to the average weight
in the subspace given by the chaotic basis states and the $M$ regular basis states
with quantum number $m$.
Here $\Pch(t)$ is still given by Eq.~\eqref{eq:normalized_preg} and $\Ncha$ is
the number of chaotic basis states
\begin{equation}
 \Ncha = N - M\mmax
 %= M (\nicefrac{1}{\heff}-\mmax)
 \;\;.
\end{equation}
The definition of the \floodweight{} $\Wm_m(t)$ is such that it
 reaches $\Wm_m=1$, if the wave packet is uniformly spread over the
chaotic sea and the $m$th regular torus in all $M$ unit cells.
(v) The periodic extension leads to
transporting regular islands, e.g.\ the $m$th torus in the $j$th unit cell is mapped to the
$(j+1)$th unit cell. The case of
non-transporting islands is discussed in Ref.~\cite{Bit2010}.

Fig.~\ref{fig:flooding_weights} shows the \floodweights{} for the designed map
with $M=144$. The qualitative behavior is similar to
Fig.~\ref{fig:flooding_weights_M1} with $M=1$, but due to the large chaotic sea
one reaches \floodweights{} closer to one.

%=======================================================================
\section{Linear Regime} \label{sec:linear_regime}
%=======================================================================

The linear increase of the \floodweights{} $\Wm_m(t)$ observed in
Figs.~\ref{fig:flooding_weights_M1}(b) and
\ref{fig:flooding_weights}(b) is a consequence of the linear increase of the
\regweights{} $\Pregm(t)$ at small times due to dynamical tunneling
\begin{eqnarray}
  {\Pregm}(t) &=& \gamma_m^{\text{ch}\to\text{reg}}\, t \;\;,
\end{eqnarray}
with the chaotic-to-regular tunneling rate $\gamma_m^{\text{ch}\to\text{reg}}$.
 From Eq.~\eqref{eq:flooding_weights_Mq} with $\Pregm \ll \Pch(t)\approx1$
follows for the \floodweight{}
\begin{eqnarray}
  \Wm_m(t) &\approx& \frac{\Ncha+M}{M} \gamma_m^{\text{ch}\to\text{reg}}\, t \;\;.
 \label{eq:approximate_flooding_tunneling_1}
\end{eqnarray}

We now express the chaotic-to-regular tunneling rate $\gamma_m^{\text{ch}\to\text{reg}}$
by the regular-to-chaotic tunneling rate $\gamma_m^{\text{reg}\to\text{ch}}=\gamma_m$.
Both rates are related to the same tunneling coupling matrix
element $\vregchm=\bra{\chireg_m}U\ket{\chich}$ by Fermi's golden rule in dimensionless form (see Appendix A in
\cite{BaeKetLoe2010})
\begin{eqnarray}
 \gamma_m^{\text{ch}\to\text{reg}} &=& 2\pi
\left\langle \left|\vregchm\right|^2\right\rangle \rho^{\text{reg}}_m
\label{eq:fermi_golden_rule_reg_ch}\\
\gamma_m=\gamma_m^{\text{reg}\to\text{ch}} &=& 2\pi
\left\langle \left|\vregchm\right|^2\right\rangle \rho^{\text{ch}}
\label{eq:fermi_golden_rule}\;\;.
\end{eqnarray}
Here $\rho^{\text{ch}}=\Ncha/(2\pi)$ is the density of chaotic basis states and
$\rho^{\text{reg}}_m=M/(2\pi)$ is the density of regular basis states
corresponding to the $m$th torus. This leads to
\begin{equation}
 \gamma_m^{\text{ch}\to\text{reg}} = \gamma_m\frac{\rho^{\text{reg}}_m}{\rho^{\text{ch}}}
 = \gamma_m\frac{M}{\Ncha} \label{eq:gammamchreg}
\end{equation}
and allows for rewriting Eq.~\eqref{eq:approximate_flooding_tunneling_1} as
\begin{eqnarray}
  \Wm_m(t) &\approx& \frac{\Ncha+M}{\Ncha} \gamma_m\, t \;\;.
 \label{eq:approximate_flooding_tunneling}
\end{eqnarray}
For $M\ll\Ncha$ this simplifies to $\Wm_m(t)\approx\gamma_m t$. In the following we use
the numerically obtained tunneling rates $\gamma_m$,
see \cite{BaeKetLoeSch2008, BaeKetLoe2010}.
For the parameters of Fig.~\ref{fig:flooding_weights_M1} we find
$\gamma_0=1.16\cdot10^{-5}, \gamma_1 = 2.94\cdot10^{-4}, \gamma_2 = 2.60 \cdot10^{-3}, \gamma_3=1.41\cdot10^{-2}$.
For the parameters of Fig.~\ref{fig:flooding_weights}, where $\heff$ is changed by $2\%$ one finds
that the tunneling rates $\gamma_m$ change by less than $15\%$.
Figures~\ref{fig:flooding_weights_M1}(b) and \ref{fig:flooding_weights}(b) show that this linear
behavior, Eq.~\eqref{eq:approximate_flooding_tunneling}, is valid almost up to the saturation regime.

Naively, one would expect that the saturation happens latest at the Heisenberg
time, which in dimensionless form is given by $\tauH=\Ncha$.
Fig.~\ref{fig:flooding_weights}(b) and in particular Fig.~\ref{fig:flooding_weights_M1}(b)
show that this is not true and that the saturation regime may be reached at much larger
times. This will be explained in Sec.~\ref{sec:time_sat}.

%=======================================================================
\section{Saturation Regime} \label{sec:saturation_regime}
%=======================================================================
In this section we study the saturation regime of the \floodweights. By introducing
an appropriate scaling parameter we find universal
properties for the asymptotic behavior of the \floodweights.

\subsection{Universal scaling of the \satval{}}

The \floodweights{} saturate on plateaus of different heights
shown in Fig.~\ref{fig:flooding_weights}. We
define the \textit{\satval{}}
\begin{equation}
 \saturm = \Braket{\Wm_m(t)}_{t} \;\;,
 \label{eq:satval_def}
\end{equation}
where $\Braket{\cdot}_t$ indicates a temporal average. Numerically one uses
a time interval, which starts at times larger than the beginning of each plateau.

\begin{figure}[t]
 \centering
 \includegraphics[width=8.5cm]{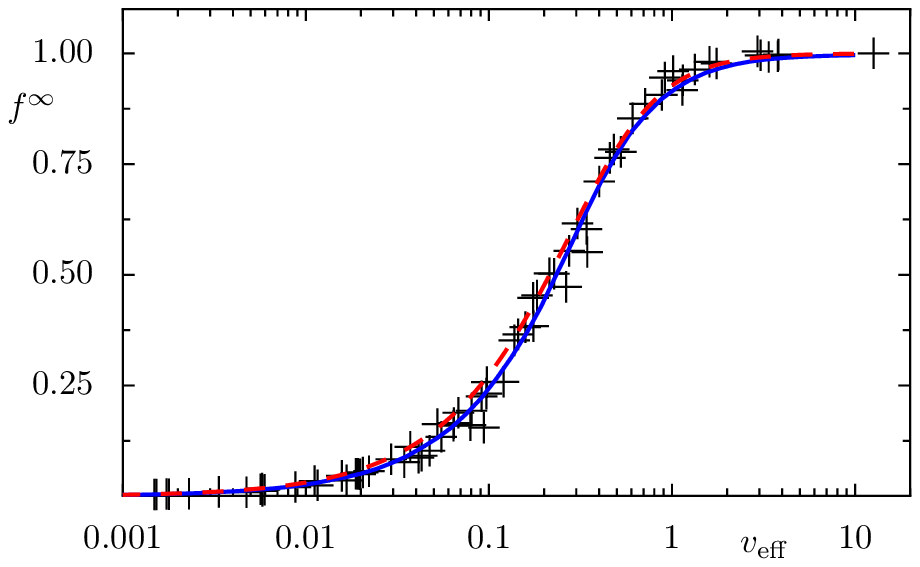}
 \caption{(Color online)
  \Satvals{} $\satur$ vs.\ effective coupling $\veff$ of the designed map for various parameters $\heff=\nicefrac{1}{10},\nicefrac{1}{20},\nicefrac{1}{30},\nicefrac{1}{40}$, $M=1,13,144,1597,17711$, and $m=1, \dots, 8$ (black crosses), random matrix prediction
 (blue solid line) and
 Eq.~\eqref{eq:saturationanalytic} (red dashed line).
 \label{fig:saturation_values}}
\end{figure}

As scaling parameter for the \satvals{}
we define an effective coupling $\veff$ between regular and chaotic basis states
\begin{equation}
 \veffm=\frac{\sqrt{\langle|\vregchm|^2\rangle}}{\Deff}\;\;.
 \label{eq:veff_def}
\end{equation}
It is obtained by rescaling the averaged tunnel coupling matrix elements $\vregchm$ with the
mean level spacing
\begin{equation}
 \Deff =\frac{2\pi}{\Ncha+M} \label{eq:Deff}
\end{equation}
in a subsystem of $\Ncha$ chaotic and $M$ regular
basis states on the $m$th regular torus.
This effective coupling $\veff$ is almost identical to the square root of the
transition parameter $\Lambda=v^2/\Delta^2$ of coupled random matrix models
for mixed systems \cite{BohTomUll1993}
with $v^2$ the variance of the perturbation and $\Delta$ the mean level spacing.
However, here the effective mean level spacing $\Deff$, Eq.~\eqref{eq:Deff},
of the subsystem
is more appropriate than the mean level spacing $\Delta=2\pi/N$.
Note, that the definition of $\Deff$ in
Eq.~\eqref{eq:Deff} is also different from the one in Ref.~\cite[Sec.~4]{BaeKetMon2007}
where $\Dcha = \nicefrac{2\pi}{\Ncha}$ is used as the effective mean level
spacing. For the
parameters used there it is irrelevant as $M\ll\Ncha$, while for $M\approx\Ncha$
we find that it is necessary to use Eq.~\eqref{eq:Deff}.

Often it is convenient to express the effective coupling $\veff$ in terms
of the numerically accessible tunneling rates $\gamma_m$
instead of the coupling matrix element $\vregchm$.
Using Fermi's golden rule, Eq.~\eqref{eq:fermi_golden_rule},
leads to
\begin{equation}
 \veffm = \frac{\Ncha + M}{2\pi\sqrt{\Ncha}}\sqrt{\gamma_m} \;\;.
 \label{eq:effective_coupling}
\end{equation}

Fig.~\ref{fig:saturation_values} shows that there is a universal
dependence of the \satvals{} $\satur$ as a function of the effective coupling $\veff$
for the designed map.
This is observed for various parameter values for
$\heff$, $M$, $\mmax$, and $m$ over several orders of magnitude in $\veff$.
Note that the transition is rather broad in  $\veff$.

\subsection{\Satval{} in terms of eigenstates}

For various applications it is convenient to express the
\satval{} in terms of properties of the eigenstates,
instead of the time evolution of wave packets.
This will include the numerical and analytical study of random matrix models
and the numerical study of the mushroom billiard.

For simplicity let us start with the case $M=1$. Furthermore we restrict
ourselves to one individual regular torus $m$ for the remainder of this section,
which is justified by the universal behavior of $\satur$ demonstrated
in Fig.~\ref{fig:saturation_values}. Then the
Hilbert space of size $N$ is spanned by an orthogonal basis given by one regular basis state
$\ket{\chireg}$ and $\Ncha=N-1$ chaotic basis states $\ket{\chich}$.

The initial wave packet $\ket{\varphi_0}$ at $t=0$
is assumed to be a random superposition of chaotic basis states $\ket{\chich_r}$
\begin{equation}
\label{eq:saturation_phi0}
\ket{\varphi_0} = \sum^\Ncha_{r=1} a_r \ue^{\ui\xi_r}\ket{\chich_r}\;\;,
\end{equation}
where $\xi_r \in [0,2\pi)$ are identical independently distributed random
variables for $r=1, \dots,\Ncha$.
The random real amplitudes $a_r$ only have to fulfill the normalization
condition $\sum_{r=1}^{\Ncha} a_r^2 = 1$ such that
\begin{equation}
\label{eq:average_amplitudes}
\langle a_r^2\rangle_{\ket{\varphi_0}}=\nicefrac{1}{\Ncha}.
\end{equation}
The expansion coefficients $c_k$ of the initial wave packet $\ket{\varphi_0}$ in the basis of
eigenstates $\ket{\psi_k}$ of $U$ are
\begin{equation}
\label{eq:saturation_ci}
c_k = \bracket{\psi_k}{\varphi_0} = \sum^\Ncha_{r=1} a_r
\ue^{\ui\xi_r}\bracket{\psi_k}{\chich_r}\;\;,
\end{equation}
for $k=1, \dots,N$. The  average over the initial wave packet $\ket{\varphi_0}$
leads to
\begin{eqnarray}
\left\langle|c_k|^2\right\rangle_{\ket{\varphi_0}} &=& \left\langle\sum^\Ncha_{r,s=1}
a_r a_s\ue^{\ui(\xi_r-\xi_s)}
\bracket{\psi_k}{\chich_r}\bracket{\chich_s}{\psi_k} \right\rangle_{\ket{\varphi_0}}\\
&=& \frac{1}{\Ncha}\sum^\Ncha_{r=1}\left|\bracket{\chich_r}{\psi_k}\right|^2\\
&=& \frac{1}{\Ncha}\left(1-\left|\bracket{\chireg}{\psi_k}\right|^2\right)\;\;,
\label{eq:saturation_average_wave_packet}
\end{eqnarray}
where in the first step the  off-diagonal terms of the double sum vanish due
to the random phases $\xi_r$ and for the
diagonal term we used Eq.~\eqref{eq:average_amplitudes}.
In the second step the completeness and orthogonality of the
regular and chaotic basis states was taken into account.

To compute the \satval{} $\satur$ we consider an ensemble of quantum maps with
associated eigenstates $\ket{\psi_k}$. Within this ensemble the regular states
and their average coupling to the chaotic states are fixed, while the chaotic
states are strongly varied. For an individual quantum map the overlap
of the time-evolved wave packet
$\ket{\varphi_t} = \sum_{k=1}^N c_k \ue^{\ui\varepsilon_k t}\ket{\psi_k}$ with the
regular basis states is according to Eq.~\eqref{eq:regular_weight} given by
\begin{eqnarray}
\Pregnonav (t)&=&\left|\bracket{\chireg}{\varphi_t}\right|^2 \\
  &=& \sum^N_{k,l=1} c_k c_l^{*}
\ue^{\ui (\varepsilon_k-\varepsilon_l)t} \bracket{\chireg}{\psi_k}
\bracket{\psi_l}{\chireg}\\
  &=& \sum^N_{k=1} |c_k|^2 \left|\bracket{\chireg}{\psi_k}\right|^2 + \Pfluct(t)
%        & & + 2\mathfrak{Re}~\left( \sum_{i>j} c_i
% c_j^{*}\ue^{2\pi\ui (E_i-E_j)t} \bracket{\chireg}{\psi_i}
% \bracket{\psi_j}{\chireg}
% \right)
\;\; ,\label{eq:saturation_flooding_value}
\end{eqnarray}
where
\begin{equation}
 \Pfluct(t) = \sum_{k\neq l} c_k c_l^{*}\ue^{\ui (\varepsilon_k-\varepsilon_l)t} \bracket{\chireg}{\psi_k} \bracket{\psi_l}{\chireg}\;\;.
\end{equation}

As in Eq.~\eqref{eq:regular_weight_av} we now perform an average over
different initial wave packets $\ket{\varphi_0}$ and perform an ensemble average
(indicated by $\Braket{\cdot}_{e}$), giving the \regweight{}
\begin{equation}
 \Preg{}(t) = \left\langle \Pregnonav (t)\right\rangle_{e,\ket{\varphi_0}}\;\;.
\end{equation}
Following Eq.~\eqref{eq:flooding_weights} and using that $\Preg{}(t)+\Pch(t)=1$
we obtain the \floodweight{}
\begin{equation}
 \Wm (t) = (\Ncha + 1)\Preg{}(t)\;\;.
\end{equation}
For the \satval{} we perform a time average according to
Eq.~\eqref{eq:satval_def}, $\satur = \Braket{\Wm (t)}_t$, and with
$\Braket{\Pfluct(t)}_t = 0$ we obtain
\begin{equation}
 \satur = (\Ncha + 1)\left\langle\sum^N_{k=1} |c_k|^2 \left|\bracket{\chireg}{\psi_k}\right|^2
 \right\rangle_{e,\ket{\varphi_0}}\;\;.
 \label{eq:satval_first}
\end{equation}

Using Eq.~\eqref{eq:saturation_average_wave_packet} we finally obtain
\begin{eqnarray}
\label{eq:saturation_single}
\satur &=& \frac{\Ncha+1}{\Ncha}\left(1- \left\langle\sum^N_{k=1}
\left|\bracket{\chireg}
{\psi_k}\right|^4\right\rangle_{e}\right)\; \; .
\end{eqnarray}
This gives the dependence of the \satval{} $\satur$ on the overlap of the eigenstates with
the regular basis states. The extreme values of $\satur$ are zero and one:
If an eigenstate of the system is identical to a regular basis state it is not
flooded at all and $\satur=0$. In contrast, for a completely flooded state,
i.e.\ $\ket{\chireg}$ has the same overlap $\nicefrac{1}{N}$ with all eigenstates $\ket{\psi_j}$,
we get $\satur=1$.

For the more general case with more than one regular basis state on the same regular torus
in different unit cells ($M\neq 1$)
we obtain with this approach the \satval{}
\begin{eqnarray}
 \satur &=& \frac{\Ncha+M}{\Ncha} \left(  1- \braket{p}_e\right)
\label{eq:saturation_Mq}\\
 p &=& \sum_{k=1}^{N} \frac{1}{M}\left(\sum_{j=1}^{M}
     \left|\bracket{\chireg_{j\phantom{k}}}{\psi_k}\right|^2\right)^2 \;\;.
\end{eqnarray}
This concludes the derivation of an expression of the
\satval{} in terms of properties of the eigenstates,
which is complementary to its definition in terms of the time evolution of wave packets,
Eq.~\eqref{eq:satval_def}.

\subsection{Random matrix modeling}
\label{sub_sec:rmt}

\begin{figure}[tb]
    \includegraphics[width=6.0cm]{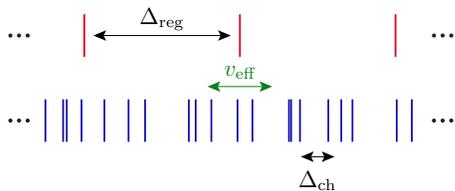}
    \caption{(Color online)
    The regular spectrum of the $m$th torus of a transporting island is equidistant with
    spacing $\Dreg$ and the chaotic spectrum is modeled by the circular orthogonal
    ensemble with mean spacing $\Dcha$. The typical regular to chaotic coupling
    is $\veff$.
    \label{fig:ladder}}
\end{figure}

Random matrix modeling has been successfully used to describe level statistics in the
context of chaos assisted tunneling, see e.g. \cite{BohTomUll1993,LeyUll1996,ZakDelBuc1998,BaeKetMon2007,TomUll1994}.
To explain the behavior of the \satvals{}
we use the model proposed in Ref.~\cite{BaeKetMon2007}
given by the time dependent Hamiltonian
\begin{equation}
H = \,
\begin{pmatrix}
 \Hreg  &  V \\
 V^{T}  & \Hcha
\end{pmatrix}\;\;.
\label{eq:hamilton_rmt}
\end{equation}
Here $\Hreg$ is a diagonal matrix with entries representing the eigenenergies
of a purely regular system, where the matrix size is given by $M\mmax$. In
the following we restrict ourselves to the study of one individual regular torus,
such that the matrix size of $\Hreg$ is $M$. In the case of transporting regular
islands the energies of these $M$ regular basis states are equidistant \cite{Sch2006},
see Fig.~\ref{fig:ladder}.
In this matrix model we choose the energy spacings such that the mean spacing $\Deff=1$
and the regular basis states are spread over an energy interval of length $N$.
% For a chosen mean spacing of $\Deff=1$ in this random matrix model
% it then follows that the regular basis states have the fixed spacing
% $\Dreg=\nicefrac{N}{M}$ and are spread over an energy interval of length $N$.
This is in contrast to the Poissonian distribution, which generically occurs if one considers many
quantized regular tori. The case of
non-transporting islands is discussed in
Ref.~\cite{Bit2010}.

For $\Hcha$ we use a diagonal matrix of size $\Ncha$, with entries representing
the eigenvalues of a purely chaotic system. Instead of using eigenvalues of a
matrix from the Gaussian orthogonal ensemble, as commonly used, we take the
eigenvalues of a matrix from the circular orthogonal ensemble \cite{Mez2007}
scaled
to the energy interval of length $N$ such that $\Dcha=\nicefrac{N}{\Ncha}$.
This ensures that we can use all eigenstates, as the mean
level spacing is constant in contrast to the semi-circle law of the Gaussian
orthogonal ensemble. Boundary effects can be neglected for sufficiently large
matrix size $N$.
The coupling matrix $V$ consists of Gaussian random variables with mean
zero and variance $(\veff \Deff)^2$ according to Eq.~\eqref{eq:Deff}, which for
the chosen energy scaling gives the variance $\veff^2$.
Numerically we used for the ratio of the submatrix sizes $\Ncha / M = 1$
and found convergence at
matrix size $N=400$ even for the largest effective couplings $\veff$. For ratios
$\Ncha / M > 1 $ convergence of $\satur$  is obtained for larger $N$ only,
e.g.\ for $\nicefrac{\Ncha}{M}=5$ a matrix size $N=10000$ is needed.

Fig.~\ref{fig:saturation_values} shows the predictions of the random matrix model for the \satvals{}, Eq.~\eqref{eq:saturation_Mq}.
They are in good agreement with the \satvals{} obtained by time evolution
of wave packets of the designed map.

\subsection{$2\times 2 $ matrix model}
\label{sec:two_by_two_model}
It turns out that a $2\times 2$ matrix model leads to the \satval{}
\begin{equation}
\saturph(\veff) = 2 \veff \arctan\frac{1}{2\veff} \label{eq:saturationanalytic}
\end{equation}
which very closely follows
the numerical \floodweights{} of the designed map
and the prediction of the random matrix model, see Fig.~\ref{fig:saturation_values}.
Equation~\eqref{eq:saturationanalytic} is the answer to the question of the
universeality \cite{Ada2002}.
It follows from a reduction of the random matrix model,
Eq.~\eqref{eq:hamilton_rmt}, to a $2\times 2$ matrix model
\begin{equation}
H = \,
\begin{pmatrix}
 \kappa  &  v \\
 v  & -\kappa
\end{pmatrix} \;\;,
\label{eq:two_by_two_matrix}
\end{equation}
where $2\kappa$ is the unperturbed spacing and $v$ the coupling matrix element
of a chaotic state $\ket{\chich}$ and the closest regular state $\ket{\chireg}$.
Their spacing $2\kappa$ is uniformly distributed in the interval $[0, 1]$ if the
number of regular and chaotic states is equal and the mean level spacing is one.
For the $2\times 2$ matrix model we therefore chose $\kappa$ to be a uniformly
distributed random variable in the interval $[0, 1/2]$.
Choosing a constant $v=\veff$ gives Eq.~\eqref{eq:saturationanalytic},
see below.
We observe that it agrees better with the numerical data than the result
obtained for a Gaussian random variable $v$.

The \satvals{} for this model can be obtained using Eq.~\eqref{eq:saturation_single}.
The overlaps $\left| \bracket{\chireg}{\psi_\pm} \right|^2$ of the eigenstates
$\ket{\psi_\pm}$ of $H$ for the
$2\times 2$ matrix are given by
\begin{equation}
 \left| \bracket{\chireg}{\psi_\pm} \right|^2 = \frac{v^2}{2}
    \left(\kappa^2 +v^2 \pm \kappa \sqrt{\kappa^2+v^2}\right)^{-1}\;\;.
\end{equation}
After some algebra Eq.~\eqref{eq:saturation_single} with $\Ncha=1$ leads for the \satval{} to
\begin{eqnarray}
\saturph &=& 2 \left(1-\left\langle \frac{\kappa^2+\frac{v^2}{2}}{\kappa^2+v^2}
\right\rangle_{e}\right) \;\;,
\end{eqnarray}
which after the ensemble average $\langle\cdot\rangle_e = 2 \int_{0}^{1/2}\ud\kappa\;\cdot\;$
leads to Eq.~\eqref{eq:saturationanalytic}.
The limits for small and strong effective coupling are
\begin{eqnarray}
\label{eq:v_approx_small}
 \saturph &\stackrel{\veff\ll 1}{=}& \pi \veff\\
 \saturph &\stackrel{\veff\gg 1}{=}& 1-1/(12\veff^2)\;\;.
\end{eqnarray}

The agreement of Eq.~\eqref{eq:saturationanalytic} with the
numerical calculations for the designed map obtained by time evolution,
see Fig.~\ref{fig:saturation_values}, is surprisingly
good even for strong coupling, where couplings to many chaotic states are relevant,
while they are not considered in the above $2\times 2$ matrix model.
This is reminiscent to the success of the Wigner surmise for level
spacing statistics which is also based on a $2\times 2$ matrix model.

\subsection{Saturation time}
\label{sec:time_sat}
We now discuss the time scale $\tplateau$ for reaching the saturation plateau.
For the $m$th regular torus we define
$\tplateaum$ by the time at which the initial linear behavior,
Eq.~\eqref{eq:approximate_flooding_tunneling},
intersects the \satval{} $\saturm$,
leading to
\begin{equation}
 \tplateaum = \frac{\Ncha \satur_m}{(\Ncha + M) \gamma_m}\;\;.
\end{equation}
By expressing $\gamma_m$ in terms of $\veff$, Eq.~\eqref{eq:effective_coupling},
and dividing by the Heisenberg time $\tauH = \Ncha+M$ we find the universal
scaling
\begin{equation}
 \frac{\tplateau}{\tauH} = \frac{\satur(\veff)}{4\pi^2\veff^2} \approx
 \frac{\saturph(\veff)}{4\pi^2\veff^2}= \frac{\arctan\frac{1}{2\veff}}{2\pi^2\veff}\;\;,
 \label{eq:tsat_analytic}
\end{equation}
where in the last step Eq.~\eqref{eq:saturationanalytic}
from the  $2\times 2$ model is used.

\begin{figure}[tb]
 \centering
 \includegraphics[width=8.5cm]{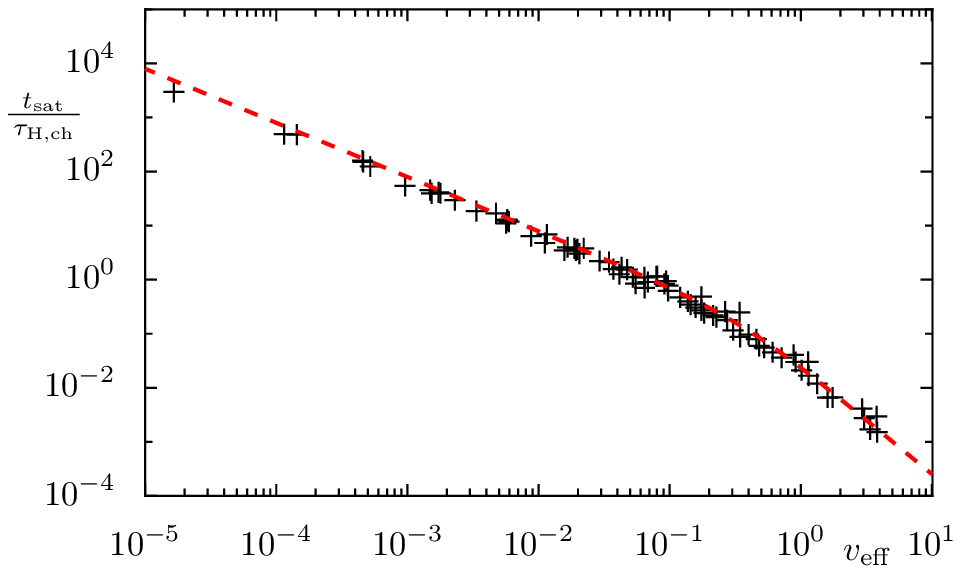}
 \caption{(Color online)
  Ratio of saturation time $\tplateau$ to Heisenberg time
 $\tauH$ vs.  effective coupling $\veff$ of the designed map for various parameters
 $\heff=\nicefrac{1}{10},\nicefrac{1}{20},\nicefrac{1}{30},\nicefrac{1}{40}$, $M=1,13,144,1597,17711$, and $m=1, \dots, 8$ (black crosses)
and Eq.~\eqref{eq:tsat_analytic} (red dashed line).
\label{fig:saturation_times}}
\end{figure}

Numerically, the saturation times are determined by the intersection time of a
fitted linear increase with the saturation plateau.
Fig.~\ref{fig:saturation_times} shows the numerical saturation times
for the designed map in comparison
to Eq.~\eqref{eq:tsat_analytic}. Very good agreement is found.
At first, it might be surprising, that the saturation time can be much larger than
the Heisenberg time. However, these large time scales arise from small
splittings of weakly coupled regular and chaotic states,
which are much smaller than the mean level spacing.

\subsection{Transition regime}
\label{sec:full_time_solution}

At the saturation time one has a transition from the linear regime to the
saturation regime. In Figs.~\ref{fig:flooding_weights_M1}(a) and
\ref{fig:flooding_weights}(a) one observes an overshooting of the
\floodweights{} beyond their asymptotic value. It can be understood by
Rabi-like oscillations that occur in the $2\times 2$ model,
Eq.~\eqref{eq:two_by_two_matrix}. For fixed coupling $v$ one obtains damped but
long-lasting oscillations. For Gaussian averaged couplings $v$ the resulting
curve resembles the numerically obtained \floodweights{}, e.g.\ as in
Fig.~\ref{fig:flooding_weights}(a) for $m=1$ \cite{Bit2010}.

The extreme case of complete flooding is studied in Appendix~\ref{app:master}.
In this case, no overshooting of the \floodweights{} is
observed, see Fig.~\ref{fig:flooding_complete}. The behavior around the
transition can be very well described by a master-equation approach.
In particular, one obtains an approximate analytical
expression for the \floodweights{},
\begin{equation}
 \Wm_m(t) = 1-\exp\left(-\frac{\Ncha+M}{\Ncha}\gamma_m^{\text{reg}\to\text{ch}} t\right)
 \label{eq:master_equation_flooding_value_solution_short}
\end{equation}
for all $m$, see Fig.~\ref{fig:flooding_complete}.

\section{Applications}
\label{sec:applications}
\subsection{Standard map}
\label{sec:flooding_standard_map}
The Chirikov standard map  \cite{Chi1979}
\begin{eqnarray}
 \label{eq-map-stdabb-tilda}
 q_{n+1} &=&  q_n + p_n \\
 p_{n+1} &=& p_n +  \frac{K}{2\pi} \sin\left(2\pi[q_n+p_n]\right)
\end{eqnarray}
is considered on the torus $[-1/2, 1/2]\times[0,1]$, i.e.\ with one unit cell
$M=1$. It arises from the kicked Hamiltonian
 \begin{equation}
 H(q,p,t) = \frac{p^2}{2} + \frac{K}{(2\pi)^2} \cos(2\pi q)
           \sum_{n\in\Z}\delta(t-n)\;\; ,
 \label{eq-ham-stdabb}
\end{equation}
with kicking strength $K$. The quantum time evolution is given by
Eq.~\eqref{eq:U_eigeneq}.

\begin{figure}[b]
 \centering
 \includegraphics[width=8.5cm]{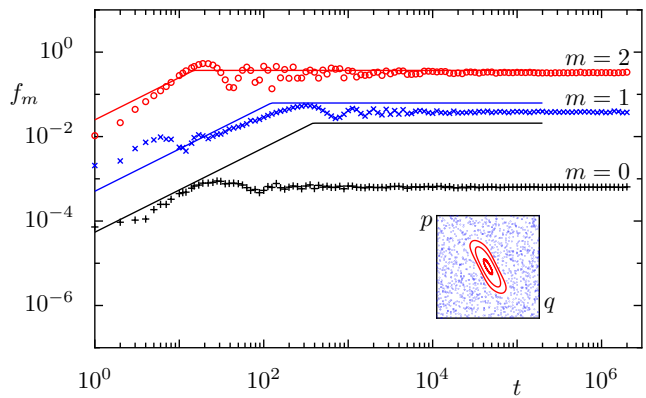}
 \caption[]{(Color online)
 \Floodweights{} $\Wm_m(t)$ of the standard map for $K=2.9$, $N=29$, and $M=1$,
 for the regular tori $m=0,1,2$ (symbols). They are compared to the prediction
 of Eqs.~\eqref{eq:approximate_flooding_tunneling} and \eqref{eq:saturationanalytic}
 (solid lines).
 \label{fig:flooding_weights_stdmap}}
\end{figure}

In Fig.~\ref{fig:flooding_weights_stdmap} the \floodweights{} are shown for
different regular tori. The qualitative behavior is the same as for the
designed map with an initial linear increase followed by a saturation plateau.
For the theoretical prediction, Eqs.~\eqref{eq:approximate_flooding_tunneling}
and \eqref{eq:saturationanalytic}, we use the numerically obtained tunneling
rates \cite{BaeKetLoeSch2008, BaeKetLoe2010}
Quantitatively we observe the largest deviation for $m=0$. We explain this
discrepancy by the less effective Bloch phase averaging, which only leads to few
avoided crossings for the standard map. This might be improved by a modified standard map
\cite{BaeKetLoeMer2011} which allows for better averaging.

As an aside we mention that the detailed behavior of the \floodweights{} for
$m=1$ can be understood by the use of $2\times 2$ matrix models \cite{Bit2010}.

\subsection{Mushroom billiard}
\label{sec:flooding_mushroom_billiard}

\begin{figure}[b]
 \centering
 \includegraphics[width=8.5cm]{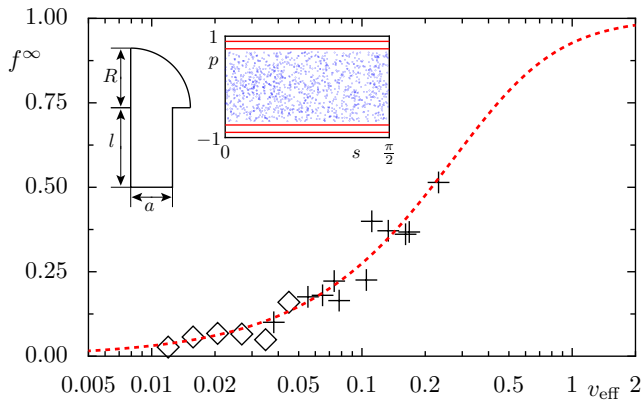}
 \caption[]{(Color online)
 \Satvals{} $\satur$ (symbols) for the mushroom billiard (inset)
  with $R=1$, $a=0.7$ vs. $\veff$
  compared to the prediction Eq.~\eqref{eq:saturationanalytic} (red dashed
 line). As explained in the text, $\veff$ is either determined analytically (diamonds) or
 numerically (crosses).
 \label{fig:saturation_values_mushroom}}
\end{figure}

The phenomenon of flooding also appears in time-independent systems with a mixed
phase space such as two-dimensional billiards. As an example we investigate the
mushroom billiard, see inset in Fig.~\ref{fig:saturation_values_mushroom}.
It is proven to have a sharply divided phase space \cite{Bun2001}.
Quantum mechanically a particle of mass $\mu$ in the
billiard can be described by the time-independent Schr\"odinger
equation, $\Delta \psi_i = E_i \psi_i$, with units $\hbar=2\mu=1$. Numerically, the
eigenfunctions $\psi_i$ are computed using the improved method of particular solutions
\cite{BetTre2005,Bet2008}.
The regular basis states are given by the eigenfunctions of the quarter circle billiard
\begin{equation}
\label{eq:chiregnm_mushroom}
 \chiregnm(\rho,\varphi) \propto \mathbf{J}_m(j_{mn}\rho) \sin(m\varphi)\;\;.
\end{equation}
in polar coordinates $(\rho,\varphi)$, where $\mathbf{J}_m$ is the $m$th
Bessel function and $j_{mn}$ is its $n$-th root.
For the determination of the \satvals{} $\satur$ it is numerically inconvenient
to use the time-evolution of wave packets. Instead we use the eigenfunctions
$\psi_i$ and determine the \satval{} $\satur$ applying Eq.~\eqref{eq:saturation_single}
to a billiard, where $\Ncha\to\infty$,
\begin{eqnarray}
\label{eq:saturation_mushroom}
\satur &=& \left(1- \sum^\infty_{i=0}\left\langle
\left|\bracket{\chiregnm}
{\psi_i}\right|^4\right\rangle_e\right)\; \; .
\end{eqnarray}
Here the ensemble average $\Braket{\cdot}_{e}$ is implemented as a variation of the
stem length $l\in[1.4, 2.3]$ for $450$ parameters with fixed $R=1$ and $a=0.7$.
Fig.~\ref{fig:saturation_values_mushroom} demonstrates the same universal scaling
of the \satval{} $\satur$ with the effective coupling $\veff$ for the mushroom
billiard as observed for the quantum map in Fig.~\ref{fig:saturation_values}.
\Satvals{} $\satur>0.6$ could not be achieved as this would require much larger
lengths of the stem, which are numerically hard to study.
For Fig.~\ref{fig:saturation_values_mushroom} we have computed the effective
coupling $\veff$ in two different ways:
(i) For small couplings we use in Eq.~\eqref{eq:effective_coupling} the
analytical result for the tunneling rates (Eq.~(8) in Ref.~\cite{BaeKetLoeRobVidHoeKuhSto2008}).
(ii) For larger couplings, where this analytical result is not accurate enough,
we determine $\veff$ from Eq.~\eqref{eq:veff_def}
using the numerically determined averaged width of
avoided crossings of the corresponding regular states under variation of the length of the
stem~\cite{BaeKetLoeRobVidHoeKuhSto2008, BaeKetLoe2010}.

\section{Summary and outlook}
\label{sec:summary}
The temporal flooding of regular tori by chaotic wave packets is
analyzed in detail.
The overlap of a wave packet started in the chaotic sea with a
regular basis state concentrated on a quantizing torus can show
Rabi-like oscillations with various amplitudes and frequencies.
We average this overlap over different initial states
and over an ensemble of quantum systems, which differ in the
chaotic region only, e.g. by varying a Bloch phase.
By a suitably defined normalization we introduce the
flooding weight, which shows universal behavior.

Initially, it increases linearly depending on the regular-to-chaotic
tunneling rate. Later it saturates and we find that the asymptotic
flooding weight shows a universal scaling with a suitably defined
effective coupling. This is found for a designed quantum map,
the standard map, and the mushroom billiard.
The universal scaling of the asymptotic flooding weight is reproduced
by a random matrix model and well described by a simple function,
Equation~\eqref{eq:saturationanalytic},
that follows from a $2 \times$ 2 matrix model.
We also find that the saturation time, at which the initial linear increase
turns into saturation, shows universal behavior that can be well
described analytically.

Beyond the present study of wave packets started in the chaotic sea,
one could study wave packets started on the regular island. They will
partially tunnel to the chaotic sea. Their asymptotic weight in
the chaotic sea, however, will depend sensitively on the initial
wave packet.
Universal behavior could be expected for the time evolution of
the regular basis states. The initial linear increase in the chaotic
region will be governed by the corresponding regular-to-chaotic
tunneling rate.
The scaling of the asymptotic weight in the chaotic sea needs to be investigated.
An experimental investigation of the consequences of flooding using
a mushroom shaped microwave cavity will be published elsewhere
\cite{BitBaeKetKuhStoe2014:p}.

\acknowledgements
We thank Steffen L\"ock for code to determine regular
basis states in the standard map and eigenfunctions
of the mushroom billiard.
We are grateful for discussions with
Martin K\"orber, Ulrich Kuhl, and Hans-J\"urgen St\"ockmann.
Furthermore, we acknowledge support by the
Deutsche Forschungsgemeinschaft within the Forschergruppe 760
``Scattering Systems with Complex Dynamics.'

%%%%%%%%%%%%%%%%%%%%%%%%%%%%%%%%%%%%%%%%%%%%%%%%%%%%%%%%%%%%%%%%%%%%%%%%%%%%%%%
\begin{appendix}

%%%%%%%%%%%%%%%%%%%%%%%%%%%%%%%%%%%%%%%%%%%%%%%%%%%%%%%%%%%%%%%%%%%%%%%%
\section{Designed map} \label{sec:amphib_map}
%%%%%%%%%%%%%%%%%%%%%%%%%%%%%%%%%%%%%%%%%%%%%%%%%%%%%%%%%%%%%%%%%%%%%%%%
% ##############################################################################
In this appendix we give the explicit definition of the
designed map and discuss some of its quantum properties.

For a kicked Hamiltonian, Eq.~\eqref{eq:hamiltonian}
a stroboscopic view after each kick gives
\begin{subequations}
\begin{eqnarray}
  q_{n+1} &=& q_n + T'(p_n) \ , \label{eq:mapQ} \\
  p_{n+1} &=& p_n - V'(q_{n+1}) \ . \label{eq:mapP}
\end{eqnarray}
\label{mapping}
\end{subequations}
We consider the mapping on the torus $[-\frac{1}{2},M-\frac{1}{2}]\times[-\frac{1}{2},\frac{1}{2}]$.
Thus the phase space extends over $M$ unit cells in $q$-direction.
The dynamics exclusively depends on the choice of the functions $T'(p_n)$ and
$V'(q_{n+1})$.

A phase space with a large regular island can be designed
\cite{HufKetOttSch2002,BaeKetMon2007} with the functions
\begin{subequations}
\label{eq:amphibtv}
\begin{eqnarray}
 v(q) &=& - k(q)\left(\frac{k(q)}{2} + x(q)\right) -\frac{rx(q)^2}{2}\;\;,\\
 t(p) &=& \left\{
              \begin{array}{ll}
                   -\frac{p}{2} + p^2  - \frac{5}{16} & \quad (p\leq0)\\
                    \frac{3p}{2} - p^2 - \frac{5}{16} & \quad (p>0)
              \end{array}
       \right.\;\;,
\end{eqnarray}
\end{subequations}
where $r=0.65$ and
\begin{subequations}
\label{eq:xm}
\begin{eqnarray}
   k(q)  &=& \left\lfloor q+\frac{1}{2}\right\rfloor\\
   x(q)  &=& q-k(q) = q-\left\lfloor q+\frac{1}{2}\right\rfloor\;\;,
\end{eqnarray}
\end{subequations}
where $\left\lfloor\cdot\right\rfloor$ denotes the floor function.
To obtain smooth functions we define
\begin{subequations}
\begin{eqnarray}
  T(p) &=& \Int_{-\infty}^{\infty} \ud z \, t(p+z) G_{\epsilon}(z), \\
  V(q) &=& \Int_{-\infty}^{\infty} \ud z \, v(q+z) G_{\epsilon}(z) .
\end{eqnarray}
\end{subequations}
where $G_{\epsilon}(z)=\exp(-z^2/2\epsilon^2)/ \sqrt{2\pi\epsilon^2}$ is a
Gaussian of width $\epsilon=0.015$. The average value of $T'(p)$ defines the
transport behavior in $q$-direction of the mapping. For the considered map
the local average velocity in the lower part of phase space ($-1/2<p<0$) is $-1$, while
in the upper part ($0<p<1/2$) it is $+1$. This leads to transport of the regular island to
the right and in the chaotic sea the diffusive transport is biased to the left.
The total phase-space velocity averages to zero.

Quantum mechanically the time evolution of a state is given by
Eq.~\eqref{eq:U_eigeneq} in terms of the unitary operator
Eq.~\eqref{eq:propagator}. The eigenstates $\ket{\psi_j}$ of this operator are
defined by Eq.~\eqref{eq:U_eigeneq_psi}.
In order to fulfill the periodicity of the classical dynamics, the quantum
states have to obey the quasi-periodicity conditions
\begin{eqnarray}
  \langle p + 1 |\psi \rangle
    &=& \ue^{- 2\pi\ui \theta_p} \langle p  |\psi \rangle  \ .
  \label{quantumperiodP}\\
  \langle q + M |\psi \rangle &=&
     \ue^{2\pi\ui \theta_q} \langle q  |\psi \rangle  \ , \label{quantumperiodQ}
\end{eqnarray}
where $\theta_q$ and $\theta_p$ are Bloch phases.
The effective Planck's constant can only be a rational number
\begin{equation}
  \heff = \frac{M}{N}   \ . \label{quanthbar}
\end{equation}
We consider the case of coprime $M$ and $N$, so that the
quantum system is not effectively reduced to less than $M$ cells.

The designed map allows for finding an analytic expression of the regular basis
states using harmonic oscillator states that are squeezed in $p$-direction,
rotated around the origin, and  shifted to the center of the regular island in the
$j$th unit cell, $q=j, p=0.25$. Neglecting the periodic boundary conditions on the torus their position
representation is given by \cite{BaeKetMon2007}
\begin{eqnarray}
   \bracket{q}{\chireg_{m,j}} &=&
   \sqrt{\frac{M}{2^m m!N}} \left( \frac{\mathfrak{Re}~c}{\pi \heffbar}
   \right)^{1/4} H_m \left( \sqrt{\frac{\mathfrak{Re}~c}{\heffbar}} (q - j) \right)
   \nonumber\\
 &&\hspace*{-0.75cm}
   \times \exp \left( - \frac{c}{2 \heffbar} (q-j)^2 +
    \frac{\ui}{4 \heffbar} (q-j/2) \right) \;\;,
   \label{eq:tiltedstate}
\end{eqnarray}
where $H_m$ is the $m$th Hermite polynomial. The complex tilting factor
$c=(\sqrt{351} - 13\,\ui)/40$
describes the orientation of the ellipse that can be derived from the
linearized map.

\section{Master equation approach for complete flooding}
\label{app:master}

\begin{figure}[tb]
 \centering
 \includegraphics[width=8.5cm]{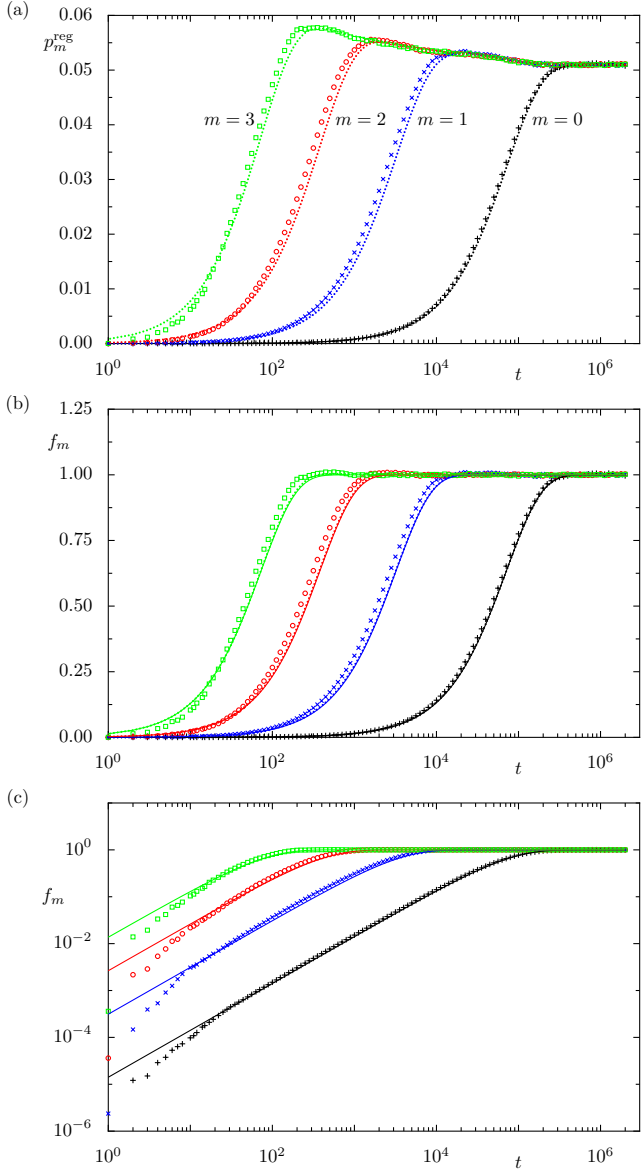}
 \caption{ a) \Regweights{} $\Pregm(t)$ of the designed map for the case of complete flooding,
    $N = 3\ 853\ 335,  M = 196\ 418, \heff=\nicefrac{1}{20}$, for $m=0,1,2,3$ (symbols)
    compared to Eq.~\eqref{eq:master} (dotted lines).
    b) Corresponding \floodweights{} $\Wm_m(t)$ and the analytical result of
    Eq.~\eqref{eq:master_equation_flooding_value_solution_short}
    (solid lines, almost indistinguishable from the dotted lines).
    c) Double-logarithmic representation of the same data as in b).
 \label{fig:flooding_complete}}
\end{figure}

In the extreme case where all regular basis states of the system are completely flooded, we can describe the time-dependent flooding by a master equation.
It considers several regular basis states and one
chaotic reservoir representing all chaotic basis states.
Their weights $\Pregm$ and $\Pch$ change due to the tunneling rates $\gamma_m^{\text{ch}\to\text{reg}}$,
Eq.~\eqref{eq:gammamchreg},
\begin{eqnarray}
 \frac{\ud}{\ud t} {\Pregm}(t) &=& \gamma_m^{\text{ch}\to\text{reg}} {\Pch}(t) -  \gamma_m^{\text{reg}\to\text{ch}}{\Pregm}(t)\\
 \frac{\ud}{\ud t} {\Pch}(t) &=& \sum_m\left(\gamma_m^{\text{reg}\to\text{ch}}{\Pregm}(t)- \gamma_m^{\text{ch}\to\text{reg}} {\Pch}(t) \right)  \;\;.
 \label{eq:master_equation_regular_weight}
\end{eqnarray}
Rephrasing these equations in matrix form gives
\begin{equation}
\label{eq:master_matrix}
 \frac{\ud}{\ud t}
 \fourvec{\Preg{0}(t)} {\vdots} {\Preg{\mmax-1}(t)} {\Pch(t)} = A
\fourvec{\Preg{0}(0)} {\vdots} {\Preg{\mmax-1}(0)} {\Pch(0)}
\end{equation}
with
\begin{equation}
A =
\left(\begin{array}{cccc}
       -\gamma_0^{\text{reg}\to\text{ch}} &  & 0 &
\gamma_0^{\text{ch}\to\text{reg}}\\
       &\ddots&  &\vdots\\
      0& &-\gamma_{\mmax-1}^{\text{reg}\to\text{ch}}
&\gamma_{\mmax-1}^{\text{ch}\to\text{reg}}\\
      \gamma_0^{\text{reg}\to\text{ch}}&\dots
&\gamma_{\mmax-1}^{\text{reg}\to\text{ch}} &
-\sum_m\gamma_m^{\text{ch}\to\text{reg}}
      \end{array}\right)\;\;.
\end{equation}
Its solution is given by
\begin{equation}
\label{eq:master}
 \fourvec{\Preg{0}(t)} {\vdots} {\Preg{\mmax-1}(t)} {\Pch(t)} =
\exp (A t)
\fourvec{\Preg{0}(0)} {\vdots} {\Preg{\mmax-1}(0)} {\Pch(0)}\;\;.
\end{equation}
For the initial conditions $\Pch(0)=1$ and $\Preg{m}(0)=0$ for all $m$,
modeling a wave packet started in the chaotic sea, good agreement
with the \regweights{} obtained for the quantum map is observed, Fig.~\ref{fig:flooding_complete}(a).
The same also holds for the \floodweights{} $\Wm_m(t)$, Eq.~\eqref{eq:flooding_weights_Mq},
see Figs.~\ref{fig:flooding_complete}(b) and (c). This example nicely demonstrates
the advantage of the \floodweights{} $\Wm_m(t)$ having the same shape for all $m$, in contrast
to the \regweights{} $\Pregm(t)$.

Considering just the subsystem of one regular state with quantum number $m$ and
the chaotic reservoir the matrix $A$ reduces to a $2\times 2$ matrix. This allows
for an analytical solution for the \regweight{} $\Pregm(t)$ from Eq.~\eqref{eq:master}
and yields Eq.~\eqref{eq:master_equation_flooding_value_solution_short}
for the \floodweight{} $\Wm_m(t)$.
This analytical result is indistinguishable in Figs.~\ref{fig:flooding_complete}(b) and (c)
from the solution of the full master equation. It describes the transition from
the initial linear increase to saturation in the case of a completely flooded
regular torus.
\end{appendix}

\end{document}